\documentclass[aps,pra,twocolumn,superscriptaddress,groupedaddress]{revtex4}
\usepackage[english]{babel}
\usepackage{amssymb}
\usepackage{cancel}
\usepackage{color,graphicx}
\usepackage{amsmath}
\usepackage{amsbsy}
\usepackage{amsthm}
\usepackage{bbm}
\usepackage{epsfig}
\usepackage{lscape}
\usepackage{float}
\usepackage{graphicx}
\usepackage{subfigure}
\usepackage{dcolumn}
\usepackage{bbm}
\usepackage{color,epstopdf}
\usepackage{amscd}
\usepackage{amsfonts}  
\usepackage[]{amsmath}
\usepackage{amssymb}    
\usepackage{mathrsfs}
\usepackage{verbatim}
\usepackage[]{cases}
\usepackage{amsmath}
\usepackage{wasysym}
\usepackage[utf8]{inputenc}
\usepackage[T1]{fontenc}
\usepackage{mathtools}

\usepackage{bbold}
\begin{document}
\title{Gravitational decoherence: a general non relativistic model }

\author{L. Asprea}
\email{lorenzo.asprea@phd.units.it}
\affiliation{Department of Physics, University of Trieste, Strada Costiera 11, 34151 Trieste, Italy}  
\affiliation{Istituto Nazionale di Fisica Nucleare, Trieste Section, Via Valerio 2, 34127 Trieste, Italy}      
\author{G. Gasbarri}
\email{g.gasbarri@uab.ca}
\affiliation{Department of Physics and Astronomy, University of Southampton, Highfield Campus, SO17 1BJ, United Kingdom}
\affiliation{Istituto Nazionale di Fisica Nucleare, Trieste Section, Via Valerio 2, 34127 Trieste, Italy}      
\date{\today}
\author{A. Bassi}
\affiliation{Department of Physics, University of Trieste, Strada Costiera 11, 34151 Trieste, Italy}
\affiliation{Istituto Nazionale di Fisica Nucleare, Trieste Section, Via Valerio 2, 34127 Trieste, Italy}
\begin{abstract}
We derive a general quantum master equation for the dynamics of a scalar bosonic particle interacting with a weak, stochastic and classical external gravitational field. The dynamics predicts decoherence in position, momentum and energy. We show how our master equation reproduces the results present in the literature by taking appropriate limits, thus explaining the apparent contradiction in their dynamical description. Our result is relevant in light of the increasing interest in the low energy quantum-gravity regime.
\end{abstract}

\maketitle

\section{Introduction}
One of the greatest predictions of general relativity is the existence of gravitational waves, which can be thought of  as small perturbations of the metric propagating through spacetime at the speed of light~\cite{maggiore,mtw,wein,carrol}. They are of fundamental interest in many branches of physics, such as cosmology, theoretical physics and astrophysics, and their recent first detection~\cite{ligo,tian,decigo,et,kagra} has opened thrilling new horizons for research and a huge effort is being put into the construction of ever more sophisticated  detectors~\cite{lisa}.\\ 
Most gravitational waves that arrive on the Earth are produced by different unresolved mechanisms and sources~\cite{allen,stoc}, and thus result in a stochastic perturbation of the flat spacetime background. Within the framework of quantum theory, this stochastic background affects the dynamics of matter propagation~\cite{aha,linet} and, when the quantum state is in a superposition, it leads to decoherence effects, as typical of noisy environments. Since quantum superpositions are very sensitive to small variations of the surrounding environment, quantum interferometers have the potential to detect a stochastic gravitational background~\cite{cow,fuentes,atint}.\\
Different models for the description of this phenomenon have been proposed~\cite{goku,breuer,sanchez,power,blencowe,ana,lamine}. However, they do not agree on the decoherence mechanism (the preferred basis and  rates) at which it takes place. 
With this work we clarify this issue. We derive a general non relativistic model of gravitational decoherence starting from the dynamics of a scalar bosonic field coupled to a weak gravitational perturbation. We show how this model recovers the results present in the literature as appropriate limiting cases.\\
\\
The paper is organized as follows. In section \ref{ii} we derive the equations of motion in Hamiltonian form for a scalar bosonic field minimally coupled to a weakly perturbed flat metric. We then specialize such equation to the non relativistic regime in section \ref{iii} and proceed with the canonical quantization of the bosonic field in the single particle sector, obtaining a Schr\"{o}dinger like equation for a test particle interacting with a weakly perturbed gravitational field.\\
In section \ref{iv} we specialize to the case of a stochastic gravitational perturbation and derive the corresponding master equation. We discuss the decoherence effect in sections \ref{v} and \ref{vi} with explicit reference to the preferred eigenbasis and characteristic decoherence time. In the same sections we show under which assumptions our master equation is able to reproduce the apparently contradictory results of~\cite{breuer,sanchez,power} and~\cite{blencowe,ana}, thus solving the preferred basis puzzle. 
\section{Hamiltonian equations of motion}\label{ii}
We first derive the equations of motion (EOM) for a scalar bosonic field minimally coupled to linearized gravity. We start from the action for the charged Klein Gordon field in curved spacetime~\cite{bd}:
\begin{equation}\label{action}
\begin{split}
S = \int d^4x \sqrt{-g}\mathcal{L}
\end{split}
\end{equation}
with the Lagrangian density:
\begin{equation}
\mathcal{L} = (c^2 g^{\mu\nu}\nabla_\mu\psi^*\nabla_\nu\psi - \frac{m^2c^4}{\hbar^2}\psi^*\psi)
\end{equation}
where $\nabla_{\mu}$ is the covariant derivative with respect to the Christoffel connection.
We write the metric as the sum of a flat background $\eta_{\mu\nu} = \text{diag}(+---)$, and a perturbation $h_{\mu\nu}$:
\begin{equation}
g_{\mu\nu} = \eta_{\mu\nu}+h_{\mu\nu}
\end{equation}
We are interested in studying the dynamics of the Klein Gordon field in presence of a weak gravitational perturbation. Therefore we perform a Taylor expansion of the action around the flat background metric and truncate the series at the first perturbative order. Thus, we obtain the effective Lagrangian $\mathcal{L}_{eff}$ acting on flat spacetime: 
\begin{equation}
\begin{split}
S = & \int d^4x \Big[ c^2(\eta^{\mu\nu}\partial_\mu\psi^{*} \partial_\nu\psi - \frac{m^2c^2}{\hbar^2}\psi^*\psi)(1+\frac{tr(h^{\mu\nu})}{2})+\\
& \quad\quad\quad - c^2h^{\mu\nu}\partial_\mu\psi^*\partial_\nu\psi + O(h^2)  \Big] \\
=&:\int d^4x (\mathcal{L}_{eff}+O(h^2))
\end{split}
\end{equation}
Note that in doing so we are implicitely restricting the analysis to the class of reference frames in which the coordinates are described by rigid rulers, 
which are negligibly affected by the gravitational perturbation. This assumption though reasonable, as measuring devices are held together by intra and iter molecular forces, is arbitrary (it may be possible that a gravitational perturbation bends a measuring device).\\
The equations of motion for the matter field are obtained (at first order in the perturbation $h_{\mu\nu}$) from the Euler-Lagrange equations:
\begin{equation}
\frac{\partial \mathcal{L}_{eff}}{\partial \psi^* } - \partial_\alpha \frac{\partial \mathcal{L}_{eff}}{\partial \partial_\alpha \psi^* } = 0
\end{equation}
and in the  harmonic gauge~\footnote{The harmonic Gauge implies translational invariance of the infinitesimal volume in the chosen coordinate system as e.g. in cartesian coordinates.} they read:
\begin{equation}\label{kg}
\begin{split}
\Big[ & -\partial_t^2 + c^2(1+h^{00})\mathbf{\nabla}^2 +2ch^{0i}\partial_t\partial_i +c^2h^{ij}\partial_i\partial_j+ \\
& -\frac{m^2c^4}{\hbar^2}(1+h^{00})+O(h^2)\Big]\psi = 0
\end{split}
\end{equation}
\\
We are interested in the description of the dynamics of a positive energy particle system in the non relativistic limit. In such a limit, the particle and antiparticle sectors are non interacting with one another, that is to say, the EOM~\eqref{kg} can be recast to a system of two uncoupled equations, one for each species sector. While this is evident and straightforward for the free case, for an interacting theory the decoupling is very complicated and achievable only perturbatively. \\
The first step is to explicitely express the field in a two component form. This can be done following the Feshbach-Villars formulation~\cite{villars}. Accordingly we define a new field:
\begin{equation}
\Psi = \left(
\begin{array}{cc}
\phi \\
\chi
\end{array}
\right)
\end{equation}
such that:
\begin{eqnarray}\label{fv}
\left\{
\begin{array}{cc}
\psi = \phi + \chi \\
i\hbar\Big(\partial_t-ch^{0i}\partial_i\Big)\psi = mc^2(\phi - \chi)
\end{array} 
\right.
\end{eqnarray}
We note that such a formulation does not allow for a probabilistic interpretation of the field $\Psi$, as the conserved charged $Q$ associated to the internal $U(1)$ symmetry ($\psi \rightarrow e^{ie}\psi$ ; $\psi^*\rightarrow e^{-ie}\psi^*$) via Noether's Theorem reads:
\begin{equation}\label{charge}
Q = 2e\: mc^2 \int d^3 x 
\left(\begin{array}{cc} \phi & \chi \end{array}\right)\sigma_3(1+\frac{tr(h^{\mu\nu})}{2}-h^{00}) \left(\begin{array}{c} \phi \\ \chi \end{array}\right)
\end{equation} 
instead of the required:
\begin{equation}
\rho = 2e\: mc^2 \int d^3 x 
\left(\begin{array}{cc} \phi & \chi \end{array}\right)\sigma_3 \left(\begin{array}{c} \phi \\ \chi \end{array}\right)
\end{equation}

We therefore apply the transformation:
\begin{eqnarray}\label{transprob}
\left\{
\begin{array}{ccc}
T &=& (1+\frac{tr(h)}{4}-\frac{h_{00}}{2})\\
\Psi &\rightarrow& T \Psi
\end{array}
\right.
\end{eqnarray}
so that, in the new representation, the squared modulus of the field can be regarded as a probability density in the non relativistic limit.\\
With the help of Eq.~\eqref{fv} and after some algebra (See Appendix A) the EOM~\eqref{kg} read:
\begin{equation}\label{eqmot}
i\hbar\partial_t \Psi = [mc^2\sigma^3 + \mathfrak{E}+ \mathcal{O}]\Psi
\end{equation}
where: 
\begin{equation}
\begin{split}
\mathfrak{E}=& \frac{mc^2}{2}h^{00}\sigma_3 -\frac{\hbar^2}{2m}(1+h^{00})\sigma_3\mathbf{\nabla}^2 -\frac{\hbar^2}{2mc}\partial_t(h^{0i})\sigma_3\partial_i\\
& -\frac{\hbar^2}{2m}h^{ij}\sigma_3\partial_i\partial_j + i\hbar c h^{0i}\partial_i - \frac{i\hbar}{2}\partial_t(\frac{tr(h^{\mu\nu})}{2}-h^{00})\\
&-\Big[\frac{\hbar^2}{4m}\nabla^2(h^{00})  - \frac{i\hbar^2}{8m}\nabla^2(tr(h^{\mu\nu}))\Big)\Big]\sigma_3 
\end{split}
\end{equation}
\begin{equation}
\begin{split}
\mathcal{O}=& \frac{imc^2}{2}h^{00}\sigma_2 -\frac{i\hbar^2}{2m}(1+h^{00})\sigma_2\mathbf{\nabla}^2 -\frac{i\hbar^2}{2mc}\partial_t(h^{0i})\sigma_2\partial_i  \\
-& \frac{i\hbar^2}{2m}h^{ij}\sigma_2\partial_i\partial_j -\Big[\frac{i\hbar^2}{4m}\nabla^2(h^{00})-\frac{i\hbar^2}{8m}\nabla^2(tr(h^{\mu\nu}))\Big]\sigma_2
\end{split}
\end{equation}
are respectively the diagonal and antidiagonal parts of the Hamiltonian $K = mc^2\sigma^3 + \mathfrak{E}+ \mathcal{O}$, and $\sigma_i$,~$i=1,2,3$ are the Pauli matrices.\\ 
\\
In the next section we will decouple the EOM to then take the non relativistic limit.
\section{Non relativistic limit and canonical quantization}\label{iii}
We want to find a representation of the two component field $\Psi$ in which
the EOM \eqref{eqmot} are diagonal. This representation can be found in non relativistic limit
following the Foldy-Wouthuysen Method~\cite{foldy}, which allows one to write perturbatively (at any order in $\frac{v}{c}$) two decoupled equations, one for each component of the field. The method is operatively characterized by the application of an appropriate transformation $U$:
\begin{equation}
\Psi  \rightarrow  \Psi^\prime = U\Psi 
\end{equation}
\begin{equation}
\begin{split}
K \rightarrow  K^\prime =& U(K-i\hbar\partial_t)U^{-1}\\
 =&  mc^2 \sigma_3+ \mathfrak{E}^{\prime} + \mathcal{O}^{\prime} +O(h^2)
\end{split}
\end{equation}
such that, in the new representation, the antidiagonal part $\mathcal{O}^\prime$ is of higher order in $\frac{v}{c}$ than the diagonal $\mathfrak{E}^\prime$. By neglecting $\mathcal{O}^\prime$ one recovers two decoupled equations. 
By performing iteratively the transformation, one can always find a representation of the two component field for which the EOM are diagonal at any desired order in $\frac{v}{c}$.\\
In our case, we have that the task is easily achieved by applying the subsequent transformations:
\begin{eqnarray}\label{trans}
\left\{
\begin{array}{ccc}
U &=& e^{- i \sigma_3 \mathcal{O} /(2mc^2)} \\
U^\prime &=& e^{-i\sigma_3 \mathcal{O^\prime} /(2mc^2)}\\
U^{\prime\prime} &=& e^{-i\sigma_3 \mathcal{O^{\prime\prime}} /(2mc^2)}
\end{array}
\right.
\end{eqnarray}
after which, with some algebra (see Appendix B), the EOM read:
\begin{equation}\label{nr}
\begin{split}
i\hbar\partial_t \Psi =& H \Psi \\
=&\Big[ mc^2(1+\frac{h^{00}}{2})\sigma_3 -\frac{\hbar^2}{2m}(1+\frac{h^{00}}{2})\mathbf{\nabla}^2\sigma_3 +\\
&\quad -\frac{\hbar^2}{2m}h^{ij}\partial_i\partial_j\sigma_3 +i\hbar c h^{0i}\partial_i +\frac{i\hbar}{2}\partial_t(h^{00})\\
&\quad -\frac{i\hbar}{4}\partial_t(tr(h^{\mu\nu})) +\frac{\hbar^2}{8m}\mathbf{\nabla}^2(tr(h^{\mu\nu}))\sigma_3 \Big] \Psi \\
& + O(c^{-4})+O(h^2_{\mu\nu})
\end{split}
\end{equation}
Note that as the transformations \eqref{trans} are generalized unitary~\cite{armin}, they preserve the conserved charge in \eqref{charge}, i.e. the probability density in the non relativistic limit.\\ 
In the non relativistic limit the EOM \eqref{nr} do not mix the two components $\phi$ and $\chi$  of the field (up to a very small correction). As we are interested in the dynamics of particles only, we restrict the analysis to the first field component $\phi$.\\ 
\\
Since the dynamics preserves the probability density, we are allowed to apply the canonical quantization prescription and impose the equal time commutation relations:
\begin{equation}
\begin{split}
[\hat{\phi}(t,\mathbf{x}),\hat{\phi}(t,\mathbf{x'})] =  & [\hat{\phi}^{\dagger}(t,\mathbf{x}),\hat{\phi}^{\dagger}(t,\mathbf{x'})] =0\\
[\hat{\phi}(t,\mathbf{x}),\hat{\phi}^{\dagger}(t,\mathbf{x'})] =& \: \delta^{3}(\mathbf{x}-\mathbf{x'})
\end{split}
\end{equation}
to obtain the EOM for the quantum field.
The equation thus obtained does not allow for the creation or annihilation of particles. We can thus safely project it onto a single particle sector to obtain the single particle Schr\"{o}dinger equation:
\begin{equation}\label{schr}
i\hbar\partial_t\vert\phi(t)\rangle = (\hat{H}_0+\hat{H}_p)\vert\phi(t)\rangle
\end{equation}
with:
\begin{equation}
\begin{split}
\hat{H}_0 =& mc^2 +\frac{\hat{\mathbf{p}}^2}{2m}\\
\hat{H}_p =& \frac{mc^2}{2}h^{00}(t,\hat{\mathbf{x}}) - \frac{\hbar^2}{8m}\lbrace h^{00}(t,\hat{\mathbf{x}}),\hat{\mathbf{p}}^2\rbrace +\frac{c}{2}\lbrace h^{0i},\hat{p}_i\rbrace \\
&-\frac{1}{4m}\lbrace h^{ij}(t,\hat{\mathbf{x}}),\hat{p}_i\hat{p}_j\rbrace +\frac{\hbar^2}{8m}\mathbf{\nabla}^2(tr[h^{\mu\nu}(t,\hat{\mathbf{x}})])+\\
&+\frac{i\hbar}{2}\partial_t(h^{00}(t,\hat{\mathbf{x}}))-\frac{i\hbar}{4}\partial_t(tr[h^{\mu\nu}(t,\hat{\mathbf{x}})])
\end{split}
\end{equation}
where $\hat{\mathbf{x}},\hat{\mathbf{p}}$ are respectively the single particle  position and the momentum operator. Note that the anticommutators between the gravitational perturbation (which is a function of the position operator) and the particle's momentum operator need to be included in the quantization prescription in order to guarantee the hermiticity of the Hamiltonian. The term $H_0$ is the standard free Hamiltonian plus an irrelevant global phase $mc^2$ that can be reabsorbed with the transformation:
\begin{equation}
\vert\phi (t)\rangle \rightarrow e^{imc^2 t/\hbar}\vert\phi (t)\rangle
\end{equation} 
The term $\hat{H}_p$ is a perturbation that encodes the interaction between the scalar bosonic particle and a weak, otherwise generic, gravitational perturbation. We note that Eq.~\eqref{schr} correctly reduces to the usual Schr\"{o}dinger equation for a particle in an external static Newtonian potential: 
\begin{equation}
\begin{split}
i\hbar\partial_t\vert\phi(t)\rangle =& \Big(\frac{\hat{\mathbf{p}}^2}{2m} - m\Phi \Big)\vert \phi(t)\rangle \\
\Phi =& -\frac{c^2 h^{00}}{2}
\end{split}
\end{equation}
if we consider the external gravitational field to be of the same form of that of the Earth.\\
\\
The generalization of Eq.~\eqref{schr} to an extended body is not an easy task, as one needs to take into account the degrees of freedom of all the elementary particles that constitute the body. However, it is rather simple to obtain the dynamics for just the center of mass if we assume that the internal degrees of freedom are frozen and cannot be excited by the gravitational perturbation as in the case of a rigid body. In such an approximation  it is convenient to define the center of mass ($\hat{\mathbf{X}}$) and relative coordinate ($\hat{\mathbf{r}}_i$) operators:
\begin{eqnarray}
\left\{
\begin{array}{ll}
\hat{\mathbf{X}}  =  \int d^3 r \: \mathbf{r} \frac{\hat{m}(\mathbf{r})}{M}     \\
\hat{\mathbf{r}}_i  =  \hat{\mathbf{x}}_i - \hat{\mathbf{X}}
\end{array}
\right.
\end{eqnarray}
and their canonical conjugates, respectively $\hat{\mathbf{P}}$ and $\hat{\mathbf{k}}_i$, where $\hat{m}(\mathbf{r})$ is the mass density operator~\footnote{Under the assumption that the rigid body consists in an ensenble of a large number N of particles, the density operator can be defined as $\hat{m}(\mathbf{x})~=~\sum\limits_i \frac{m_{i}}{(2\pi\hbar)^{3}}\int d \mathbf{q}\: e^{-\frac{i}{\hbar}(\mathbf{x}-\mathbf{\hat{x}_i})\cdot{\mathbf{q}}}$, where $m_{i}$ and $\hat{x}_{i}$ are the mass and the position operator of the i-th particle.} and $M~=~\int d^3 r \; \hat{m}(\mathbf{r})$ is the total mass. Upon tracing out the relative degrees of freedom, the Hamiltonian for the center of mass of a rigid body reads:
\begin{equation}\label{rigid}
\begin{split}
\hat{H} =&Mc^2+\frac{\hat{\mathbf{P}}^2}{2M}  + \int d^3 r\:  h^{00}(\mathbf{r},t) m(\hat{\mathbf{X}}+\mathbf{r})c^2 +\\
-&\int d^3 r \: h^{00}(\mathbf{r},t)\frac{\lbrace m(\mathbf{r+\hat{\mathbf{X}}}),\hat{\mathbf{P}}^2\rbrace}{8M^2}\\
+& c \int d^3r \: h^{0i}(\mathbf{r},t)\frac{\lbrace m(\mathbf{r}+\hat{\mathbf{X}}), \hat{P}_i\rbrace}{2M} \\
-& \int d^3 r \: h^{ij}(\mathbf{r},t)\frac{\lbrace m(\mathbf{r+\hat{\mathbf{X}}}),\hat{P}^i\hat{P}^j\rbrace}{4M^2}\\
+&\frac{\hbar^2 c^2}{8M} \int d^3 r \: \nabla^2(tr[h^{\mu\nu}(\mathbf{r},t)]) \frac{m(\hat{\mathbf{X}}+\mathbf{r})}{M} +\\
+&\frac{i\hbar c^2}{2}\int d^3 r \: \partial_t\Big( h^{00}(\mathbf{r},t) - \frac{1}{2}tr(h^{\mu\nu}(\mathbf{r},t)) \Big)\frac{m(\hat{\mathbf{X}}+\mathbf{r})}{M}
\end{split}
\end{equation}
Eq.~\eqref{rigid} was derived following the work of~\cite{giulio} where, however, the authors only consider the special case with $h^{0i}~=~h^{ij}=0$.
\\
In the next section we will specialize to the case of a (weak) stochastic gravitational background.
\section{Stochastic gravitational perturbation: single particle master equation}\label{iv}
The motivation to consider a stochastic weak gravitational perturbation is given by the interest towards Stochastic Semi-classical Gravity (an attempt to self-consistently describe the back-reaction of the quantum stress-energy fluctuations on the gravitational field, without having to invoke the quantization of the latter; see for example~\cite{hu} and \cite{stoch2} for a review and further references), and by the interest in a stochastic gravitational background (see for instance~\cite{allen,stoc}), which we have already briefly introduced in Section I.
\\
\\
If the metric is random Eq.~\eqref{schr} becomes a stochastic differential equation. As a consequence the predictions are given by taking the stochastic average over the random gravitational field. We then need to specify its stochastic properties.\\
We assume the noise to be Gaussian and with zero mean. The first assumption is justified by the law of large numbers, while the second by our choice of taking from the very beginning the Minkowski spacetime as the background spacetime around which the metric fluctuates. For the sake of simplicity, we also assume the different components of the metric fluctuation to be uncorrelated. This means that the noise is fully characterized by:
\begin{equation}\label{stocprop}
\begin{split}
\mathbb{E} [ h_{\mu\nu}(\mathbf{x},t)] =& 0\\
\mathbb{E} [h_{\mu\nu}(\mathbf{x},t)h_{\mu\nu}(\mathbf{y},s)] =& \alpha^2f_{\mu\nu}(\mathbf{x},\mathbf{y};t,s)
\end{split}
\end{equation}
\\
where $\mathbb{E}[\, \cdot \,]$ denotes the stochastic average and $\alpha$ represents the strength of the gravitational fluctuations.
he two point correlation function $f(\mathbf{x},\mathbf{y};t,s)$ is a real function of order one, i.e. $0 \le\vert f_{\mu\nu}(\mathbf{x},\mathbf{y};t,s)\vert\le 1$.\\
\\
We move to the density operator formalism~\footnote{We note that that of state vectors (and Schr\"{o}dinger equation) is not the most appropriate formalism to adopt for the description of a quantum stochastic process in most experimental situation, as it does not allow one to describe statistical mixtures of quantum states.}:
\begin{equation}
\hat{\Omega}(t)=\vert\phi(t)\rangle\langle\phi(t)\vert
\end{equation}
As the only characterization of the noise is given by the stochastic average (Eq.~\eqref{stocprop}), we study the dynamics of the averaged operator:
\begin{equation}
\hat{\rho}(t)=\mathbb{E} [\hat{\Omega}(t)]
\end{equation}
Let us consider the von Neumann equation for the averaged density matrix :
\begin{equation}
\begin{split}\label{von}
\partial_{t}\hat{\rho}(t)=& -\frac{i}{\hbar}\Big[ \hat{H}_0(t),\hat{\rho}(t)\Big]-\frac{i}{\hbar}\mathbb{E}\Big[ [\hat{H}_p(t),\hat{\Omega}(t)]\Big]\\
 \equiv & \mathbb{E}\Big[ \mathfrak{L}[\hat{\Omega(t)}]\Big]
\end{split}
\end{equation}
where $\mathfrak{L}[\,\cdot\,]$ denotes the Liouville superoperator. Equation~\eqref{von} is in general difficult to tackle, because of the stochastic average, but it can be solved perturbatively by means of  the cumulant expansion~\cite{vank} (see Appendix C). With the further help of the Gaussianity, zero mean, uncorrelation of different components, we can rewrite Eq.~\eqref{von}  in Fourier space~\footnote{Our choice for the Fourier transform is: $$f(\mathbf{x}) = \frac{1}{(\sqrt{2\pi}\hbar)^{3}}\int d^3q\:\tilde{f}(\mathbf{q}) e^{i\mathbf{q}\cdot\mathbf{x}/\hbar} $$} as:
\begin{widetext}
\begin{equation}\label{nonmarkov}
\begin{split}
\partial_t \hat{\rho} = & -\frac{i}{\hbar}[\hat{H}_0,\hat{\rho}(t)]+\\
&-\frac{\alpha^2}{\hbar^8}\int \frac{d^3q\:d^3q^\prime}{(2\pi)^{3}}\int_{0}^{t}dt_{1}\:\tilde{f}^{00}(\mathbf{q},\mathbf{q^\prime};t,t_1)\frac{m(\mathbf{q})m(\mathbf{q^\prime})}{M^2}\Big[ e^{i\mathbf{q}\cdot\hat{\mathbf{X}}/ \hbar}(\frac{\hat{P}^2}{4M}+\frac{Mc^2}{2}),[e^{i\mathbf{q^\prime}\cdot\hat{\mathbf{X}}_{t_1}/\hbar}(\frac{\hat{P}^2}{4M}+\frac{Mc^2}{2}),\hat{\rho}(t)\Big]\Big] + \\
&  -\frac{\alpha^2c^2}{\hbar^8}\int\frac{d^3q\:d^3q^\prime}{(2\pi)^{3}}\int_{0}^{t}dt_{1}\:\tilde{f}^{0i}(\mathbf{q},\mathbf{q^\prime};t,t_1)\frac{m(\mathbf{q})m(\mathbf{q^\prime})}{M^2}\Big[e^{i\mathbf{q}\cdot\hat{\mathbf{X}}/ \hbar}\hat{P}_i,\Big[e^{i\mathbf{q^\prime}\cdot\hat{\mathbf{X}}_{t_1}/\hbar}\hat{P}_i,\hat{\rho}(t)\Big]\Big]+\\
& -\frac{\alpha^2}{\hbar^8}\int \frac{d^3q\:d^3q^\prime}{(2\pi)^{3}}\int_{0}^{t}dt_{1} \:\tilde{f}^{ij}(\mathbf{q},\mathbf{q^\prime};t,t_1)\frac{m(\mathbf{q})m(\mathbf{q^\prime})}{M^2}\Big[e^{i\mathbf{q}\cdot\hat{\mathbf{X}}/ \hbar}\frac{\hat{P}_i\hat{P}_j}{2M},\Big[e^{i\mathbf{q^\prime}\cdot\hat{\mathbf{X}}_{t_1}/\hbar}\frac{\hat{P}_i\hat{P}_j}{2M},\hat{\rho}(t)\Big]\Big]+\\
&-\frac{\alpha^2}{\hbar^8}\int \frac{d^3q\:d^3q^\prime}{(2\pi)^{3}}\int_{0}^{t}dt_{1}\:\tilde{f}^\mu_\mu(\mathbf{q},\mathbf{q^\prime};t,t_1)\frac{\mathbf{q}^2\mathbf{q^\prime}^2}{64M^2}\frac{m(\mathbf{q})m(\mathbf{q^\prime})}{M^2}\Big[e^{i\mathbf{q}\cdot\hat{\mathbf{X}}/ \hbar},\Big[e^{i\mathbf{q^\prime}\cdot\hat{\mathbf{X}}_{t_1}/\hbar},\hat{\rho}(t)\Big]\Big]+\\ & -\frac{\alpha^2}{16\hbar^4}\int \frac{d^3q\:d^3q^\prime}{(2\pi)^{3}}\int_{0}^{t}dt_{1}\:\partial_t\partial_{t_1}\tilde{f}^{\mu}_{\mu}(\mathbf{q},\mathbf{q^\prime};t,t_1)\frac{m(\mathbf{q})m(\mathbf{q^\prime})}{M^2}\Big[e^{i\mathbf{q}\cdot\hat{\mathbf{X}}/ \hbar},\Big[e^{i\mathbf{q^\prime}\cdot\hat{\mathbf{X}}_{t_1}/\hbar},\hat{\rho}(t)\Big]\Big]+ O(t\alpha^4\tau_c^3)\end{split}
\end{equation}
\end{widetext}
where $\hat{x}_{t_1} = e^{i\hat{H}_0t_{1}}\hat{x}e^{-i\hat{H}_0t_{1}}$. Note that the above equation becomes exact if $[\hat{H}_0,\hat{H}_p]=0$ (see Appendix D). The above equation describes the dynamics of the rigid body's center of mass is in the presence of an external weak, stochastic gravitational field (with the further assumptions made in this section), and constitutes the main result of this paper.\\
In the following we will not consider the effect on the dynamics due to the derivatives of the metric perturbation, as in typical experimental situations~\cite{ligo,tian,decigo,et,kagra} they are negligible and in any case they would not add any further informative content to the analysis. This means that we neglect the last line of Eq.~\eqref{nonmarkov}.\\
\\
We now restrict our analysis to the Markovian case, i.e. we assume the noise to be delta correlated in time:
\begin{equation}
f^{\mu\nu}(\mathbf{x},\mathbf{y};t,s) = j^{\mu\nu}(\mathbf{x},\mathbf{y};t)\delta(t-s)
\end{equation} 
A further reasonable assumption, motivated by the homogeneity of spacetime itself, is that of translational invariance of the two point correlation function:
\begin{equation}\label{correlation}
f^{\mu\nu}(\mathbf{x},\mathbf{y};t,s) = \lambda u^{\mu\nu}(\mathbf{x}-\mathbf{y})\delta(t-s)
\end{equation} 
where the factor $\lambda$ is in principle a generic coefficient with the dimension of a time. Note that the white noise assumption makes physical sense only if the correlation time ($\tau_{c}$) of the gravitational fluctuations is much smaller than the free dynamics' characteristic time  ($\tau_{free}$), or in the case where the contribution to the dynamics due to the gravitational perturbation is not affected by the free evolution dynamics, i.e. the operators describing the perturbation commute with the free dynamics operator $\hat{H}_0$ (See Appendix D). In such cases, as a first approximation, we can take $\lambda$ to be:
\begin{equation}\label{eq:lambda}
\lambda = \min(\tau_c \: ,\: t)
\end{equation}
Note that this choice does not affect the generality of the analysis as we leave $u^{\mu\nu}(\mathbf{x}-\mathbf{y})$ unspecified.\\
In such a regime Eq.~\eqref{nonmarkov} is exact and it is easy to show that it reduces to:
\begin{widetext}
\begin{equation}\label{ext}
\begin{split}
\partial_t \hat{\rho} = & -\frac{i}{\hbar}[\hat{H}_0,\hat{\rho}(t)]\\
 & -\frac{\alpha^2 \lambda c^4}{4(2\pi)^{3/2}\hbar^5 }\int d^3q\:\tilde{u}^{00}(\mathbf{q})m^2(\mathbf{q})\Big[e^{i\mathbf{q}\cdot\hat{\mathbf{X}}/\hbar},\Big[e^{-i\mathbf{q}\cdot\hat{\mathbf{X}}/\hbar},\hat{\rho}(t)\Big]\Big]  \\
& -\frac{\alpha^2 \lambda}{(2\pi)^{3/2}\hbar^5 }\int d^3q \:\tilde{u}^{00}(\mathbf{q})\frac{m^2(\mathbf{q})}{M^2}\:\Big[\Big\lbrace e^{i\mathbf{q}\cdot\hat{\mathbf{X}}/ \hbar},\frac{\hat{\mathbf{P}}^2}{4M}\Big\rbrace ,\Big[\Big\lbrace e^{-i\mathbf{q}\cdot\hat{\mathbf{X}}/\hbar},\frac{\hat{\mathbf{P}}^2}{4M}\Big\rbrace ,\hat{\rho}(t)\Big]\Big]\\
& -\frac{\alpha^2 \lambda c^2}{2(2\pi)^{3/2}\hbar^5 }\int d^3q \:\tilde{u}^{00}(\mathbf{q})\frac{m^2(\mathbf{q})}{M}\:\Big[e^{i\mathbf{q}\cdot\hat{\mathbf{X}}/ \hbar},\Big[\Big\lbrace e^{-i\mathbf{q}\cdot\hat{\mathbf{X}}/\hbar},\frac{\hat{\mathbf{P}}^2}{4M}\Big\rbrace,\hat{\rho}(t)\Big]\Big]\\
& -\frac{\alpha^2 \lambda c^2}{2(2\pi)^{3/2}\hbar^5 }\int d^3q \:\tilde{u}^{00}(\mathbf{q})\frac{m^2(\mathbf{q})}{M}\:\Big[\Big\lbrace e^{i\mathbf{q}\cdot\hat{\mathbf{X}}/ \hbar},\frac{\hat{\mathbf{P}}^2}{4M}\Big\rbrace,\Big[e^{-i\mathbf{q}\cdot\hat{\mathbf{X}}/\hbar},\hat{\rho}(t)\Big]\Big]\\
&  -\frac{\alpha^2\lambda c^2}{(2\pi)^{3/2}\hbar^5 }\int d^3q \:\tilde{u}^{0i}(\mathbf{q})\frac{m^2(\mathbf{q})}{4M^2}\:\Big[\Big\lbrace e^{i\mathbf{q}\cdot\hat{\mathbf{X}}/ \hbar},\hat{P}_i\big\rbrace,\Big[\Big\lbrace e^{-i\mathbf{q}\cdot\hat{\mathbf{X}}/\hbar},\hat{P}_i\Big\rbrace ,\hat{\rho}(t)\Big]\Big]\\
& -\frac{\alpha^2\lambda}{(2\pi)^{3/2}\hbar^5 }\int d^3q \:\tilde{u}^{ij}(\mathbf{q})\frac{m^2(\mathbf{q})}{M^2}\:\Big[\Big\lbrace e^{i\mathbf{q}\cdot\hat{\mathbf{X}}/ \hbar},\frac{\hat{P}_i\hat{P}_j}{4M}\Big\rbrace,\Big[\Big\lbrace e^{-i\mathbf{q}\cdot\hat{\mathbf{X}}/\hbar},\frac{\hat{P}_i\hat{P}_j}{4M}\Big\rbrace,\hat{\rho}(t)\Big]\Big]
\end{split}
\end{equation}
\end{widetext}
Eq.~\eqref{ext} describes decoherence  both in position and in momentum, as it contains double commutators of functions of the position, momentum and free kinetic energy operators respectively with the averaged density matrix. In particular, we immediately recognize the term in the second line of Eq.~\eqref{ext} to give decoherence in position, that in the third line might give decoherence in energy (in the regime in which $\frac{\mathbf{q}\cdot\hat{\mathbf{X}}}{\hbar}$ is small), and that in the sixth line decoherence in momentum (in the same regime).\\
    \\
In the next section we will investigate under which conditions Eq.~\eqref{ext} reduces the different models of gravitational decoherence present in the literature.
\section{Decoherence in the position eigenbasis}\label{v}
In this section we specialize Eq.~\eqref{ext} to the regime in which the dominant contribution to the decoherence effect is in the position eigenbasis. This can be done under the following assumptions:
\begin{eqnarray}
\left\{
\begin{array}{ll}
h^{00}\gtrsim h^{0i}\\
h^{00}\gtrsim h^{ij} \\
\Delta E \ll Mc^2\:(1-u^{00}(\Delta \mathbf{x})) 
\end{array}
\right.
\end{eqnarray}
where $\Delta \mathbf{x}$ and $\Delta E$ are the quantum coherences of the system, respectively the position and energy ($E = \frac{\mathbf{P}^2}{2M}$). It is then easy to show that the leading contribution to Eq.~\eqref{ext} is:
\begin{equation}\label{sgp}
\begin{split}
&\partial_t \hat{\rho} =  -\frac{i}{\hbar}[\hat{H},\hat{\rho}(t)]+\\
&\: -\frac{\alpha^2 \tau_c c^4}{(2\pi)^{3/2}\hbar^5}\int d^3q \:\tilde{u}^{00}(\mathbf{q})m^2(\mathbf{q})\:\Big[e^{i\mathbf{q}\cdot\hat{\mathbf{X}}/\hbar},\Big[e^{-i\mathbf{q}\cdot\hat{\mathbf{X}}/\hbar},\hat{\rho}(t)\Big]\Big]\\
&\:+O(h^{\mu i})+O(\Delta E)
\end{split}
\end{equation}
where we have safely replaced $\lambda=\tau_c$. The above equation describes decoherence in the position eigenbasis as the Lindblad operator is a funciton of the position operator. It is actually of the same form of the Gallis-Fleming master equation~\cite{gallis}, which describes the decoherence induced on a particle by collisions with a surrounding thermal gas, allowing for a collisional interpretation of the result. \\
To compare with the previous literature on gravitational deocherence, we must further characterize the spatial correlation function of the noise and the mass density distribution. We start by considering the model proposed by Blencowe~\cite{blencowe}. In order to recover an analogous master equation we must assume 
the noise to be delta correlated in space:
\begin{equation}
u^{00}(\mathbf{x}-\mathbf{x^\prime}) = l^3\delta^3(\mathbf{x}-\mathbf{x^\prime})
\end{equation}
where $l$ is a generic coefficient with the dimension of a length. Under this assumptions Eq.~\eqref{ext}, represented in the position eigenbasis, in fact becomes:
\begin{equation}\label{blencowe}
\begin{split}
&\partial_t\rho(\mathbf{x},\mathbf{x^\prime};t) = \quad \frac{i\hbar}{2M}(\nabla^2_x-\nabla^2_{x^{\prime}})\rho(\mathbf{x},\mathbf{x^\prime};t)+\\
& -\frac{(\alpha^{00})^2\tau_c c^4l^3}{4\hbar^2}\int d^3r\:\Big(m(\mathbf{r}-\mathbf{x})-m(\mathbf{r}-\mathbf{x^\prime})\Big)^2\rho(\mathbf{x},\mathbf{x^\prime};t)\\
&+O(h^{\mu i})
\end{split}
\end{equation}
which has the same form of the master equation obtained in~\cite{blencowe}, and describes decoherence in position. The different rate is due to the different treatment of the gravitational noise: Blencowe considers a quantum bosonic thermic bath whose correlation functions can not be reproduced by our classical description of the noise. If we further take the mass density function to be a Gaussian:
\begin{equation}\label{mass}
m(\mathbf{r}) = \frac{m}{(\sqrt{2\pi}R)^3}e^{-\mathbf{r}^2/(2R^2)}
\end{equation}
as it is done in the same work, Eq.~\eqref{blencowe} then reads:
\begin{equation}
\begin{split}
&\partial_t\rho(\mathbf{x},\mathbf{x^\prime};t) = \quad \frac{i\hbar}{2M}(\nabla^2_x-\nabla^2_{x^{\prime}})\rho(\mathbf{x},\mathbf{x^\prime};t)+\\
& -\frac{\alpha^2M^2\tau_c c^4l^3}{4(\sqrt{\pi})^3\hbar^2R^3}\Big(1-e^{-\frac{(\mathbf{x}-\mathbf{x^\prime})^2}{4R^2}}\Big)\rho(\mathbf{x},\mathbf{x^\prime};t)\\
&+O(h^{\mu i})
\end{split}
\end{equation}
To recover the results obtained by Sanchez Gomez~\cite{sanchez}, we instead first need to take the mass density function to be ponitlike:
\begin{equation}\label{eq:massdelta}
m(\mathbf{r}) = M \delta^3(\mathbf{r})
\end{equation}
as in~\cite{sanchez}, and then to assume the spatial correlation function to be Gaussian: 
\begin{equation}\label{corr}
\tilde{u}^{00}(\mathbf{q}-\mathbf{q^\prime}) = L^{3}\hbar^3\delta(\mathbf{q}-\mathbf{q^\prime})e^{-\hbar^2\mathbf{q}^2L^2/2}
\end{equation}
where $L$ is the correlation length of the noise. With this choice for the spatial correlation functions it is natural to assume
\begin{equation}
\tau_c = \frac{L}{c}    
\end{equation}
as it is the only time scale of the system left, and Eq.~\eqref{sgp} represented in the position basis reduces to:
\begin{equation}\label{sg}
\begin{split}
\partial_t\rho(\mathbf{x},\mathbf{x^\prime};t) = &\quad \frac{i\hbar}{2m}(\nabla^2_x-\nabla^2_{x^{\prime}})\rho(\mathbf{x},\mathbf{x^\prime};t)+\\
&+\frac{2\alpha^2 m^2c^3L}{\hbar^2}\Big( e^{-\frac{(\mathbf{x}-\mathbf{x^\prime})^2}{2L^2}}-1 \Big)\rho(\mathbf{x},\mathbf{x^\prime};t)
\end{split}
\end{equation}
and exactly recovers Sanchez Gomez's result.\\
A very similar equation was also obtained by Power and Percival~\cite{power}. Our model is able to qualitative recover the shape of the master equation, but not the specific rate which depends of the fourth power of the noise's strength, being the anlysis in \cite{power} at higher order in the gravitational perturbation expansion.\\
\\
\\
In the next section we will describe under which assumptions our model is able to describe decoherence in the momentum and energy eigenbasis thus encompassing the results of Breuer et al.~\cite{breuer} that predict gravitational decoherence to occur in the energy eigenbasis.
\section{Decoherence in the momentum eigenbasis}\label{vi}
In this section we specialize Eq.~\eqref{ext} to the regime in which the dominant contribution to the decoherence effect is in the momentum or energy eigenbasis. This is the case when we can approximate:
\begin{equation}
e^{i\mathbf{q}\cdot\hat{\mathbf{X}}/\hbar}\sim \hat{\mathbb{1}}
\end{equation}
i.e. in the case of small $\mathbf{q}$. In this case Eq.~\eqref{ext} reduces to:
\begin{equation}\label{momentum}
\begin{split}
\partial_t \hat{\rho} = & -\frac{i}{\hbar}[\hat{H},\hat{\rho}(t)]\\
-& \frac{\alpha^2 \lambda}{(2\pi)^{3/2}\hbar^5 }\int d^3q \:\tilde{u}^{00}(\mathbf{q})\frac{m^2(\mathbf{q})}{M^2}\:\Big[\frac{\hat{\mathbf{P}}^2}{2M},\Big[\frac{\hat{\mathbf{P}}^2}{2M},\hat{\rho}(t)\Big]\Big]\\
-&  \frac{\alpha^2\lambda c^2}{(2\pi)^{3/2}\hbar^5 }\int d^3q \:\tilde{u}^{0i}(\mathbf{q})\frac{m^2(\mathbf{q})}{M^2}\:\Big[\hat{P}_i,\Big[\hat{P}_i,\hat{\rho}(t)\Big]\Big]\\
-& \frac{\alpha^2 \lambda}{(2\pi)^{3/2}\hbar^5 }\int d^3q \:\tilde{u}^{ij}(\mathbf{q})\frac{m^2(\mathbf{q})}{M^2}\:\Big[\frac{\hat{P}_i\hat{P}_j}{2M},\Big[\frac{\hat{P}_i\hat{P}_j}{2M},\hat{\rho}(t)\Big]\Big]
\end{split}
\end{equation}
In order to recover the results of Breuer et al.~\cite{breuer}, the following hierarchy of the gravitational fluctuation must be verified:
\begin{eqnarray}
\left\{
\begin{array}{cc}
h^{ij}\gg h^{0i}\\
h^{ij}\gg h^{00}
\end{array}
\right.
\end{eqnarray}
and the spatial correlation functions are chosen to be:
\begin{equation}
\tilde{u}^{ij}(\mathbf{q}-\mathbf{q^\prime}) = \delta^{ij}L^3\hbar^3\delta(\mathbf{q}-\mathbf{q^\prime})e^{-\hbar^2\mathbf{q}^2L^2/2}
\end{equation}
Also in this case it is natural to choose $\tau_c = L/c$. We also assume the mass density distribution to describe a pointlike particle as in Eq.~\eqref{eq:massdelta}.\\
Under these assumptions Eq.~\eqref{momentum} in fact reduces to:
\begin{equation}\label{breuer}
\partial_t \hat{\rho} = -\frac{i}{\hbar}[\hat{H},\hat{\rho}(t)]-\frac{\alpha^2 \lambda}{\hbar^2}
\Big[\frac{\hat{\mathbf{P}}^2}{2M},\Big[\frac{\hat{\mathbf{P}}^2}{2M},\hat{\rho}(t)\Big]\Big]
\end{equation}
Eq.~\eqref{breuer} is indeed the same as the one obtained by Breuer et al. with the identification:
\begin{equation}
\alpha^2\lambda = \frac{T_c}{2}
\end{equation} 
where $T_c$ is the spatially averaged correlation time of the noise present in the same paper~\footnote{In the work of Breuer et al. the symbol used for the spatially averaged correlation time is $\tau_c$. It was here changed to $T_c$ in order to avoid any confusion with our own notation.}.\\
With the same assumptions we are also able to reproduce the shape of the master equation derived by Anastopoulos and Hu \cite{ana}, but not the exact rate. As in the case of the Blencowe model, this is due to their quantum treatment of the gravitational noise.
\section{Conclusions}\label{vii}
In this paper we have derived a general model of decoherence for a non relativistic quantum particle interacting with a weak stochastic gravitational perturbation. We have specialized such an equation to the Markovian limit under some further reasonable assumptions on the stochastic properties of the gravitational noise motivated by simplicity arguments and cosmological models and observations.  \\
We have extended our model to the description of the center of mass of a rigid extended body, which is a more realistic and experimentally interesting scenario.\\
Our Markovian master equation predicts decoherence in position, momentum and energy as it contains, among other terms, double commutators of functions of the position, momentum and free kinetic energy operators with the averaged density matrix.\\
We were able to succesfuly recover other results present in the literature as appropriate limiting cases of our general master equation. 
\section*{Acknowledgments}
LA  thanks L. Curcuraci, J.L. Gaona Reyes and C.I. Jones for the helpful and inspiring discussions. The authors acknowledge financial support form the EU Horizon 2020 research and innovation program under Grant Agreement No. 766900 [TEQ]. LA and AB thank the University of Trieste and INFN. GG thanks the Leverhulme Trust [RPG- 2016-046]. AB thanks the COST action QTSpace. 
\appendix
\section{ Feshbach Villars formalism }
Here we provide explicit calculation for the derivation of Eq.~\eqref{eqmot}.\\
Let us first rewrite Eq.~\eqref{kg} as:
\begin{equation}\label{fv2}
\begin{split}
(i\hbar\partial_t - i\hbar c h^{0i}\partial_i)^2 \psi &= \Big[ \hbar^2c\partial_t(h^{0i})\partial_i - \hbar^2c^2(1+h^{00})\mathbf{\nabla}^2 +\\
 &-\hbar^2 c^2h^{ij}\partial_i\partial_j+m^2c^4(1+h^{00})\Big]\psi +\\
 &+O(h^2)
\end{split}
\end{equation}
and the system of Eq.~\eqref{fv} as
\begin{eqnarray}
\left\{
\begin{array}{cc}
i\hbar(\partial_t-ch^{0i}\partial_i)\psi + mc^2\psi = 2mc^2\phi \\
i\hbar(\partial_t-ch^{0i}\partial_i)\psi - mc^2\psi = - 2mc^2\chi 
\end{array}
\right.
\end{eqnarray}
Casting Eq.~\eqref{fv2} in the above system we get :
\begin{equation}
\begin{split}
i\hbar(\partial_t-ch^{0i}\partial_i)\phi =& \frac{mc^2}{2}(\phi-\chi) +\\
&+\frac{m^2c^4}{2mc^2}(1+h^{00})(\phi+\chi)+\\
& -\frac{\hbar^2}{2m}(1+h^{00})\mathbf{\nabla}^2(\phi+\chi)+\\
&-\frac{\hbar^2}{2m}h^{ij}\partial_i\partial_j(\phi+\chi)+\\
&+\frac{\hbar^2}{2mc}\partial_t(h^{0i})\partial_i(\phi+\chi)
\end{split}
\end{equation}
\begin{equation}
\begin{split}
i\hbar(\partial_t-ch^{0i}\partial_i)\chi =& -\frac{mc^2}{2}(\phi-\chi) +\\
&-\frac{m^2c^4}{2mc^2}(1+h^{00})(\phi+\chi) +\\
&+\frac{\hbar^2}{2m}(1+h^{00})\mathbf{\nabla}^2(\phi+\chi)+\\
&+\frac{\hbar^2}{2m}h^{ij}\partial_i\partial_j(\phi+\chi) +\\
&-\frac{\hbar^2}{2mc}\partial_t(h^{0i})\partial_i(\phi+\chi)
\end{split}
\end{equation}
Recalling now that  $\Psi = \left(\begin{array}{c} \phi \\ \chi \end{array}\right)$ and exploiting the Pauli matrices, the system reduces to:
\begin{equation}
\begin{split}
i\hbar\partial_t \Psi =&  \Big[mc^2 \sigma_3 +\frac{mc^2}{2}h^{00}[\sigma_3+i\sigma_2] +i\hbar c h^{0i}\partial_i \\
-&\frac{\hbar^2}{2m}(1+h^{00})[\sigma_3+i\sigma_2]\mathbf{\nabla}^2 + \\
-&\frac{\hbar^2}{2mc}\partial_t(h^{0i})[\sigma_3+i\sigma_2]\partial_i+\\
-&\frac{\hbar^2}{2m}h^{ij}[\sigma_3+i\sigma_2]\partial_i\partial_j\Big]\Psi\\
=&: \mathfrak{H}\Psi
\end{split}
\end{equation}
Upon applying the transformation~\eqref{transprob},
the EOM transform as:
\begin{equation}
\mathfrak{H}\rightarrow K := T \mathfrak{H}T^{-1}+ i\hbar T\partial_t(T^{-1})
\end{equation}
and read exactly as Eq.~\eqref{eqmot} of the main text.
\section{Foldy Wouthuysen method}
Here we illustrate the Fouldy Wouthuysen method applied to Eq.~\eqref{eqmot}.
Let us consider the transformations: 
\begin{equation}
K \rightarrow  K^\prime = U(K-i\hbar\partial_t)U^{-1}
\end{equation}
and specialize $U$ to Eq.~\eqref{trans}, i.e.
\begin{equation}
U = e^{- i \sigma_3 \mathcal{O} /(2mc^2)} =: e^{iS}
\end{equation}
With the help of the BCH identity:
\begin{equation}
\begin{split}
K^\prime = & e^{iS} (K-i\hbar\partial_t) e^{-iS} = K + i [S,K] +\frac{i^2}{2!}[ S[S,K]] +\\
&+ \frac{i^3}{3!} [S[ S[S,K]]] + ...\\
&+\hbar(-\dot{S}-\frac{i}{2}[S,\dot{S}]+\frac{1}{6}[S,[S,\dot{S}]]+...)
\end{split}
\end{equation}
Recalling that:
\begin{equation}
K = mc^2\sigma_3 + \mathfrak{E} + \mathcal{O}
\end{equation}
and noticing that: 
\begin{eqnarray}
\lbrack\sigma_3,\mathfrak{E}\rbrack &=& 0 \\
\lbrace\sigma_3 , \mathcal{O} \rbrace &=& 0 \\
\lbrack\sigma_3 \mathcal{O},\sigma_3\rbrack &=& -2\mathcal{O}\\
\lbrack\sigma_3 \mathcal{O}, \mathfrak{E}\rbrack &=& \sigma_3 \lbrack \mathcal{O},\mathfrak{E}\rbrack\\
\lbrack\sigma_3 \mathcal{O} ,\mathcal{O} \rbrack &=& 2 \sigma_3 \mathcal{O}^2
\end{eqnarray}
it is not difficult to check that:
\begin{equation}
K^\prime = mc^2\sigma_3 +\mathfrak{E}^{\prime} + \mathcal{O}^\prime 
\end{equation}
where:
\begin{equation}
\begin{split}
\mathfrak{E}^\prime =& \mathfrak{E} + \sigma_3(\frac{\mathcal{O}^2}{2mc^2}-\frac{\mathcal{O}^4}{8m^3c^6})-\frac{i}{8m^2c^4}[\mathcal{O},\dot{\mathcal{O}}]\\
&-\frac{1}{8m^2c^4}[\mathcal{O},[\mathcal{O},\mathfrak{E}]]+...
\end{split}
\end{equation}
\begin{equation}
\mathcal{O}^\prime = \frac{1}{2mc^2}\sigma_3[\mathcal{O},\mathfrak{E}]-\frac{\mathcal{O}^3}{3m^2c^4}+\frac{i}{2mc^2}\sigma_3\dot{\mathcal{O}}+...
\end{equation}
We note that $\mathcal{O}^\prime$ is of order $c^{-1}$, meaning that we need to perform a further transformation if we want non trivial diagonal EOM. 
The transformation that we perform is:
\begin{equation}
U^\prime = e^{-i\sigma_3\mathcal{O}^\prime/(2mc^2)} 
\end{equation}
after which the Hamiltonian reads:
\begin{equation}
K^{\prime\prime} = mc^2\sigma_3 + \mathfrak{E}^\prime + \mathcal{O}^{\prime\prime}+...
\end{equation}
with:
\begin{equation}
\mathcal{O}^{\prime\prime}=\frac{\sigma_3}{2mc^2}[\mathcal{O}^\prime,\mathfrak{E}^\prime]+\frac{i}{2mc^2}\sigma_3\dot{\mathcal{O}^\prime}+...
\end{equation}
As $\mathcal{O}^{\prime\prime}\sim O(\frac{v^3}{c^3})$ we need to perform a final transformation:
\begin{equation}
U^{\prime\prime} = e^{-i\sigma_3\mathcal{O}^{\prime\prime}/(2mc^2)} 
\end{equation}
Finally the Hamiltonian reads:
\begin{equation}
H:= K^{\prime\prime\prime} = mc^2\sigma_3 + \mathfrak{E}^\prime +O(c^{-4})
\end{equation}
It is easy to note that the only (other than $\mathfrak{E}$) contribution to $\mathfrak{E}^\prime$ at the desired order is:
\begin{equation}
\begin{split}
\frac{\sigma_3}{2mc^2}\mathcal{O}^2 =& \frac{\sigma_3}{2mc^2} \lbrace \frac{imc^2}{2}h^{00}\sigma_2 \:,\: -\frac{i\hbar^2}{2m}\mathbf{\nabla}^2\sigma_2\rbrace+\\
& + O(h^2)+O(c^{-4})\\
=& \frac{\hbar^2}{4m}(h^{00}\mathbf{\nabla}^2+\mathbf{\nabla}^2(h^{00}))\sigma_3 + O(h_{\mu\nu}^2)+O(c^{-4})
\end{split}
\end{equation} 
so that the Hamiltonian becomes:
\begin{equation}
\begin{split}
H =& mc^2(1+\frac{h^{00}}{2})\sigma_3 -\frac{\hbar^2}{2m}(1+\frac{h^{00}}{2})\mathbf{\nabla}^2\sigma_3 +\\
&\quad -\frac{\hbar^2}{2m}h^{ij}\partial_i\partial_j\sigma_3 +i\hbar c h^{0i}\partial_i +\frac{i\hbar}{2}\partial_t(h^{00})\\
&\quad -\frac{i\hbar}{4}\partial_t(tr(h^{\mu\nu})) +\frac{\hbar^2}{8m}\mathbf{\nabla}^2(tr(h^{\mu\nu}))\sigma_3 +\\
& + O(c^{-4})+O(h^2_{\mu\nu})
\end{split}
\end{equation}
as in Eq.~\eqref{nr} of the main text.
\section{Cumulant expansion}
In this section we derive Eqs. (\ref{nonmarkov}, \ref{ext}) with the help of the Cumulant Expasion method \cite{vank}. We start by giving a brief presentation of the method, which is generally speaking a very useful tool for the solution of Stochastic Differential Equations (SDEs), to eventually apply it to our specific cases of interest.\\
\\
Let us consider the generic multiplicative SDE:
\begin{equation}\label{diffstoc}
\dot{\hat{\Omega}}(t) = [\mathcal{A}+\alpha \mathcal{B}(t)]\hat{\Omega}(t)
\end{equation}
where $\hat{\Omega}$ is a density operator, $\mathcal{A}$ a constant superoperator, $\mathcal{B}(t)$ a random superoperator with finite correlation time $\tau_{c}$, and $\alpha$ the parameter measuring the magnitude of the fluctuations. Of this form is Eqs. \eqref{von} of the main text. Our goal in this section will be to solve such an equation.\\
In the interaction picture, Eq.~\eqref{diffstoc} reads:
\begin{eqnarray}
\hat{\Omega}(t)&=& e^{t\mathcal{A}}\tilde{\hat{\Omega}}(t)\\
\dot{\hat{\tilde{\Omega}}}(t) &=& \alpha e^{-t\mathcal{A}} \mathcal{B}(t)e^{t\mathcal{A}}\hat{\tilde{\Omega}}\equiv \alpha \tilde{\mathcal{B}}(t)\hat{\tilde{\Omega}}(t)
\end{eqnarray}
Its formal solution is:
\begin{equation}\label{formal}
\hat{\tilde{\Omega}}(t)= T[e^{\alpha\int_{0}^{t}\tilde{\mathcal{B}}(s)ds}]\hat{\tilde{\Omega}}(0)
\end{equation}
Note that Eq.~\eqref{formal} represents the solution only for a given realization of the random process, while in experiments one is typically interested into averaged effects. We therefore consider the averaged differential equation:
\begin{equation}\label{diffstoc2}
\partial_t\hat{\tilde{\rho}}(t) = \mathbb{E}[\alpha \tilde{\mathcal{B}}(t)\hat{\Omega}(t)]
\end{equation}
where we recall $\hat{\rho}=\mathbb{E}[\hat{\Omega}]$. Its formal solution reads:
\begin{equation}\label{formal2}
\hat{\tilde{\rho}}(t)= \mathbb{E}\Bigg[T\left[ e^{\alpha\int_{0}^{t}\tilde{\mathcal{B}}(s)ds}\right]\Bigg]\hat{\tilde{\Omega}}(0)
\end{equation}
which is in most cases though of any practical use. In order to  alternatively solve the averaged dynamics we note that, as $\mathcal{B}(t)$ is indeed a random variable, by definition it follows that $\mathbb{E}\left[ e^{\alpha\int_{0}^{t}\tilde{\mathcal{B}}(s)ds}\right]$ is a moment generating function. We can then apply the standard cumulant expansion method (for all practical purposes a series expansion of the exponential, for more details see chapter III.4 of \cite{vank}). With such a method, we intend to find the generator of the averaged dynamics governing the statistical operator $\hat{\rho}(t)$, i.e. the non stochastic superoperator $\mathcal{G}$ such that:
\begin{equation}
\partial_t\hat{\tilde{\rho}}(t)= \mathcal{G}(t)\hat{\tilde{\rho}}(t)
\end{equation}
Upon applying the cumulant expansion to Eq.~\eqref{formal2}, we obtain:
\begin{equation}\label{eq:cumulantsol}
\begin{split}
\hat{\tilde{\rho}}(t) =T\Bigg[ e^{\frac{\alpha^2}{2}\int_{0}^{t}\int_{0}^{t_{1}}dt_{1}dt_{2}\mathbb{E}\left[\tilde{\mathcal{B}}(t_{1})\tilde{\mathcal{B}}(t_{2})\right]}\Bigg] \hat{\tilde{\Omega}}(0)
\end{split}
\end{equation}
Upon applying the cumulant expansion to Eq.~\eqref{formal}, we obtain:
\begin{equation}\label{eq:cumulant}
\begin{split}
\hat{\tilde{\rho}}(t) =& T\Big[ \exp\Big\lbrace \alpha\int_{0}^{t}dt_{1}\langle\langle\tilde{\mathcal{B}}(t_{1})\rangle\rangle\\
&+  \frac{\alpha^2}{2}\int_{0}^{t}dt_{1}dt_{2}\langle\langle\tilde{\mathcal{B}}(t_{1})\tilde{\mathcal{B}}(t_{2})\rangle\rangle\\
&+...\\
&+\frac{\alpha^m}{m!}\int_{0}^{t}dt_{1}...dt_{m}\langle\langle\tilde{\mathcal{B}}(t_{1})...\tilde{\mathcal{B}}(t_{m})\rangle\rangle +...\Big\rbrace\Big]\tilde{\Omega}(0) 
\end{split}
\end{equation}
where $\langle\langle\: \tilde{\mathcal{B}}(t_1)...\tilde{\mathcal{B}}(t_m)\:\rangle\rangle$ denotes the $m$th cumulant. 
Note that each term in the cumulant expansion is of order $O(\alpha^m\tau_c^{m-1}t)$. In the case of a Gaussian and white noise however, all terms with $m$ greater than 2 vanish \cite{isserlis}. Furthermore, In most physically interesting cases (like for Eq. ~(\ref{von}), where the the stochastic noise has zero mean), the dominant contribution to Eq.~\eqref{eq:cumulant} is given by the second order term. 
Eq.~\eqref{eq:cumulant} therefore reads:
Eq.~\eqref{eq:cumulantsol} is simpler than Eq.~\eqref{eq:cumulant}, but we are still not able to straightforwardly extract the generator of the dynamics $\mathcal{G}$ from it. In order to do so, we make use of the Disentangling Theorem \cite{disentangling} as it is presented in \cite{vankampen2}.
We therefore define a generic non stochastic time dependent superoperator $\mathcal{K}(t)$ and the relative evolution superoperator:
\begin{equation}\label{eq:v}
\mathcal{V}(t,t_1)= T\Bigg[e^{\int_{t_1}^t dt'\: \mathcal{K}(t')} \Bigg]    
\end{equation}
With the help of $\mathcal{V}(t,t_1)$ we can define a new representation for $\hat{\tilde{\Omega}}(t)$ and $\tilde{\mathcal{B}}(t)$ as:
\begin{align}
\hat{\tilde{\Omega}}(t)=&\mathcal{V}(t,0)\hat{\tilde{\Omega}}^{k}(t)    \\
\tilde{\mathcal{B}}^k(t)=&\mathcal{V}(t,0)^{-1}\tilde{\mathcal{B}}(t)\mathcal{V}(t,0)
\end{align}
so that Eq.~\eqref{eq:cumulantsol} reads:
\begin{equation}\label{eq:cumulantsol2}
\begin{split}
\hat{\tilde{\rho}}(t)& = T \Bigg[ e^{\int_0^t \mathcal{K} (t_1) dt_1}\Bigg]\cdot\\
&\cdot T\Bigg[ e^{\frac{\alpha^2}{2}\int_{0}^{t}\int_{0}^{t_{1}}dt_{1}dt_{2}\mathbb{E}\Big[\tilde{\mathcal{B}}^k(t_{1})\tilde{\mathcal{B}}^k(t_{2})\Big]-\int_0^t \mathcal{K}^k (t_1)dt_1 } \Bigg] \hat{\tilde{\Omega}}(0)
\end{split}
\end{equation}
We then conveniently choose $\mathcal{K}(t)$ such that: 
\begin{equation}\label{eq:k}
\mathcal{K}^k(t_1)= \frac{\alpha^2}{2} \int_0^{t_1}dt_2 \mathbb{E}[\tilde{\mathcal{B}}^k(t_1)\tilde{\mathcal{B}}^k(t_2)]
\end{equation}
and we are able to cancel the terms of order $\alpha^2\tau_c$ in the second factor of Eq. \eqref{eq:cumulantsol2} \cite{vankampen2}. Note that the superoperator $\mathcal{K}(t)$ is to be intended as a time local superoperator, i.e. even if defined through the integral expression in eq.~\eqref{eq:k}, the time ordering in eq.~\eqref{eq:cumulantsol2} will order the whole operator $K(t)$ only according to the time $t$. Furthermore, note that the expression for $\mathcal{K}$ is implicit:
\begin{equation}
\mathcal{K}(t_1)= \frac{\alpha^2}{2}\int_0^{t_1}dt_2 \mathbb{E}[\tilde{\mathcal{B}}(t_1)\mathcal{V}(t_1,t_2)\tilde{\mathcal{B}}(t_2)\mathcal{V}(t_1,t_2)^{-1}]
\end{equation}
as on the r.h.s. $\mathcal{V}(t_1,t_2)$ depends on $\mathcal{K}$ itself. Noticing that $\mathcal{K}$ is of $O(\alpha^2\tau_c)$, we perform a perturbative expansion in $\alpha \tau_c$ ($\mathcal{K}=\mathcal{K}_1+\mathcal{K}_2+\mathcal{K}_3+...$) in order to obtain its explicit expression. The first term ($\mathcal{K}_1$) is obtained by neglecting the action of $\mathcal{V}(t_1,t_2)$ on $\tilde{\mathcal{B}}(t_2)$ in Eq.~\eqref{eq:k} so that:
\begin{equation}
\mathcal{K}_1(t)= \frac{\alpha^2}{2}\int_0^{t_1}dt_2\mathbb{E}[\tilde{\mathcal{B}}(t_1)\tilde{\mathcal{B}}(t_2)]   
\end{equation}
The next term ($\mathcal{K}_2$) is of order $O(\alpha^4\tau_c^2)$, and is obtained upon plugging the above expression in Eq.~\eqref{eq:v}:
\begin{equation}
\begin{split}
\mathcal{K}_2(t_1)=\int_0^{t_1}dt_2 \mathbb{E}\Bigg[&\tilde{\mathcal{B}}(t_1)T\Big[e^{\int_{t_2}^{t_1}dt'\mathcal{K}_1(t')}\Big]\tilde{\mathcal{B}}(t_2)\cdot \\
&\cdot T\Big[e^{-\int_{t_2}^{t_1}dt'\mathcal{K}_1(t')}\Big]\Bigg]    
\end{split}
\end{equation}
Higher order terms can be obtained in a similar fashion. This procedure can be repeated for the other terms of the cumulant expansion, so to obtain a disentangled expression at the desired order in $\alpha$ and $\tau_c$, see \cite{vankampen2} for the explicit construction in a more general case. It follows that at $O(\alpha^2\tau_c)$ Eq.~\eqref{eq:cumulantsol} reads:
\begin{equation}\label{eq:cumulantsol3}
\begin{split}
\hat{\tilde{\rho}}(t) & =\Bigg(  T\Bigg[ e^{\int_0^t \mathcal{K}(t_1) dt_1+ O(\alpha^4\tau_c^2)}\Bigg]\cdot\\
&\cdot T\Bigg[ e^{\frac{\alpha^2}{2}\int_{0}^{t}\int_{0}^{t_{1}}dt_{1}dt_{2}\mathbb{E}\left[\tilde{\mathcal{B}}^k(t_{1})\tilde{\mathcal{B}}^k(t_{2})\right]-\int_0^t \mathcal{K}^k(t_1)dt_1 } \Bigg]\Bigg)\hat{\tilde{\Omega}}(0) 
\end{split}
\end{equation}
Eq.~\eqref{eq:cumulantsol3} is the formal solution of the differential equation:
\begin{equation}
\partial_t\hat{\tilde{\rho}}(t) = \frac{\alpha^2}{2}\int_{0}^{t}dt'\mathbb{E}[\tilde{\mathcal{B}}(t)\tilde{\mathcal{B}}(t')]\hat{\tilde{\rho}}(t)+O(\alpha^4\tau_c^3t)
\end{equation}
which in the original representation reads:
\begin{equation}\label{cumulantfinal}
\begin{split}
\partial_t\hat{\rho}(t) =& \Big(\mathcal{A}+\frac{\alpha^2}{2}\int_{0}^{t}dt'\mathbb{E}[\mathcal{B}(t)e^{\mathcal{A}(t-t')}\mathcal{B}(0)\hat{e}^{-A(t-t')}]\Big)\hat{\rho}(t)\\
&+O(\alpha^4\tau_c^3t)     
\end{split}
\end{equation}
In order to apply this result to Eq. (\ref{von}), the mapping from Eq.~\eqref{cumulantfinal} is given by:
\begin{eqnarray}\label{mapping}
\left\{
\begin{array}{cc}
\mathcal{A}= -\frac{i}{\hbar}\left(\hat{H}_{0,L} - \hat{H}_{0,R}\right)\\
\alpha \mathcal{B}= -\frac{i}{\hbar}\left(\hat{H}_{p,L}-\hat{H}_{p,R}\right)
\end{array} 
\right.
\end{eqnarray}
where the subscripts $L$ and $R$ denote the fact that the operator is acting respectively on the left and on the right of the density operator $\hat{\Omega}$ (i.e. $\mathcal{A}_L\mathcal{A}_R\hat{\Omega}=\mathcal{A}\hat{\Omega}\mathcal{A}$). The final result (at order $\alpha^2\tau_c$) is:
\begin{widetext}
\begin{equation}\label{eq:bos}
\partial_t \hat{\rho} = -\frac{i}{\hbar}[\hat{H}_0,\hat{\rho}(t)]  -\frac{1}{\hbar^2}\int_{0}^{t}dt_{1}\mathbb{E}\Bigg[\Big[\hat{H}_p(t),\Big[e^{i\hat{H}_0(t_1-t)}\hat{H}_p(t_1)e^{-i\hat{H}_0(t_1-t)},\hat{\rho}(t)\Big]\Big]\Bigg] 
\end{equation}
\end{widetext}
precisely as in Eq. (\ref{nonmarkov}) of the main text.
\section{Recovering Markovian master equation}
In this section we specialize Eq. \eqref{cumulantfinal} to interesting limiting cases. We also recover the Markovian master equation (\ref{ext}) of the main text.\\
\\
We start by considering the special case in which the stochastic superoperator $\mathcal{B}$ can be factorized as:
\begin{equation}\label{bhf}
\mathcal{B}(t)= \mathfrak{h}_{i}(t)\mathcal{F}^{i}(t)    
\end{equation}
where $\mathfrak{h}_i(t)$ is a (collection of) stochastic process(es) and $\mathcal{F}^{i}(t)$ a non stochastic superoperator.
Of this form is in fact the stochastic superoperator $\mathcal{B}$ defined in Eq. (\ref{mapping}) through the explicit expressions of Eq. (\ref{rigid}) of the main text.\\
We then notice that Eq. \eqref{cumulantfinal} becomes exact if $[\mathcal{A},\mathcal{B}]=0$. In this case in fact $[\tilde{\mathcal{B}}(t),\tilde{\mathcal{B}}(t_1)]=0=[\tilde{\mathcal{B}}^k(t)\tilde{\mathcal{B}}^k(t_1)]$, so that $\mathcal{K}=\mathcal{K}_1$, and the factor inside the second time ordering in Eq. \eqref{cumulantfinal} vanishes. It follows that Eq. \eqref{cumulantfinal} can be further simplified as: 
\begin{equation}\label{cumulantcomm1}
\begin{split}
\partial_t\hat{\rho}(t) = & \Big(\mathcal{A}+\frac{\alpha^2}{2}\int_{0}^{t}dt'D_{ij}(t,t')\mathcal{F}^{i}(0)\mathcal{F}^{j}(0)\Big)\hat{\rho}(t)    
\end{split}
\end{equation}
where $D_{ij}(t,t')=\mathbb{E}[\mathfrak{h}_{i}(t)\mathfrak{h}_{j}(t-t')]$ is the time correlation function of the noise. As a very rough approximation we take the time correlation function to be an Heaviside theta function \footnote{A more physically meaningful result can be obtained by taking the correlation function as an exponential decay}:
\begin{equation}
D_{ij}(t)=\sigma_{ij}\Theta(t-\tau_c)
\end{equation}
where $\tau_c$ is the correlation time of the noise, and $\sigma_{ij}$ depends on the explicit form of $\mathcal{B}$. In this case
Eq. \eqref{cumulantcomm1} reads:
\begin{equation}\label{cumulantcomm2}
\partial_t\hat{\rho}(t) = \Big(\mathcal{A}+\frac{\alpha^2\lambda\sigma_{ij}}{2}\mathcal{F}^{i}(0)\mathcal{F}^{j}(0)\Big)\hat{\rho}(t)
\end{equation}
where $\lambda = \min (t,\tau_c)$.\\
\\
A different interesting scenario to consider is when the Markovian limit of Eq. \eqref{cumulantfinal} can be taken, i.e. when the correlation time ($\tau_{c}$) of the noise is much smaller than the  characteristic time ($\tau_{\text{\tiny free}}$) of the free dynamics, and the limit $\tau_c/\tau_{\text{\tiny free}}\rightarrow 0$ can be taken.
I this limit the action of $e^{\mathcal{A}(t_1-t)}$ on $\mathcal{B}$ (and more generally of any of the evolution superoperators employed in the derivation of Eq. \eqref{cumulantfinal}) will vanish to zero and the equation  Eq.~\eqref{cumulantfinal} reads \footnote{Note that we have safely replaced the upper limit of integration $t$ with $\infty$, as the integrand vanishes anyway \cite{vank}.}:
\begin{equation}\label{cumulantm}
\partial_t\hat{\rho}(t) = \Big(\mathcal{A}+\frac{\alpha^2}{2}\int_{0}^{\infty}dt'D_{ij}(t,t')\mathcal{F}^{i}(0)\mathcal{F}^{j}(0)\Big)\hat{\rho}(t)
\end{equation}
This equation can be further simplified noticing that in the limit $\tau_c/\tau_{\text{\tiny free}}\rightarrow 0$ the time correlation function is naturally replaced by a Dirac delta function:
\begin{equation}
D_{ij}(t)=\sigma_{ij}\delta (t-\tau_c)    
\end{equation}
and Eq. \eqref{cumulantm} consequently reads:
\begin{equation}\label{cumulantm2}
\partial_t\hat{\rho}(t) = \Big(\mathcal{A}+\frac{\alpha^2\tau_c\sigma_{ij}}{2}\mathcal{F}^{i}(0)\mathcal{F}^{j}(0)\Big)\hat{\rho}(t)
\end{equation}
As a final remark, note that the factor $\tau_c$ in the above equation can be safely replaced with $\lambda$, as the error made lies within the boundaries of the validity of the Markovian approximation.\\
Upon subsituting the explicit expression for $\mathfrak{h}_i(t)$ and $\mathcal{F}^i(t)$ according to Eqs.~(\ref{bhf},~\ref{mapping}, \ref{rigid}) and given the stochastic properties of the noise (Eqs.~(\ref{stocprop},~\ref{correlation})), we recover Eq. (\ref{ext}) of the main text.
\bibliographystyle{unsrt}
\bibliography{biblio}

\end{document}